\documentclass[prd]{revtex4}

\usepackage{amsbsy}
\usepackage{amssymb}
\usepackage{amsmath}
\usepackage{graphicx}
\usepackage{hyperref}
\usepackage[usenames]{color}
\usepackage{rotating}

\newcommand{\preup}[1]{\,^{#1}\!}

\def\beq{\begin{equation}}
\def\eeq{\end{equation}}
\def\be{\begin{equation}}
\def\ee{\end{equation}}

\newcommand{\ep}{\rho}
\newcommand{\half}{{\frac{1}{2}}}
\newcommand{\beqa}{\begin{eqnarray}}
\newcommand{\eeqa}{\end{eqnarray}}
\newcommand{\ba}{\begin{array}}
\newcommand{\ea}{\end{array}}
\newcommand{\bea}{\begin{eqnarray}}
\newcommand{\eea}{\end{eqnarray}}
\newcommand{\bean}{\begin{eqnarray*}}
\newcommand{\eean}{\end{eqnarray*}}
\newcommand{\bpm}{\begin{pmatrix}}
\newcommand{\epm}{\end{pmatrix}}
\newcommand{\muc}{\check\mu}

\def\x{\mathrm{x}}
\def\n{\mathrm{n}}
\def\p{\mathrm{p}}

\def\H1{{H_1}}

\def\a00{{{\cal A}_0^0}}
\def\b00{{{\cal B}_0^0}}
\def\c00{{{\cal C}_0^0}}
\def\D00{{{\cal D}_0^0}}

\begin{document}

\title{Tidal deformations of neutron stars: The role of stratification and elasticity}

\author{A.J. Penner$^1$, N. Andersson$^1$, L. Samuelsson$^{2,3}$, I. Hawke$^1$ \&  D.I. Jones$^1$}

\affiliation{$^1$Mathematical Sciences, University of Southampton, Southampton SO17 1BJ, UK}

\affiliation{$^2$ Department of Physics, Ume\aa\ University, SE-901 87 Ume\aa, Sweden}

\affiliation{$^3$  Nordita, Roslagstullsbacken 23, SE-106 91 Stockholm, Sweden}

\begin{abstract}
We discuss the response of  neutron stars to the tidal interaction in a compact binary system, as encoded in the Love number associated with the 
induced deformation. 
This problem is of interest for gravitational-wave astronomy as there may be 
a detectable imprint on the signal from the late stages of  binary coalescence.
 Previous work has focussed on simple barotropic neutron star models,
providing an understanding of the role of the stellar compactness and overall density profile. We add realism to the discussion by 
developing the framework required to model stars with varying composition and an elastic crust. 
These effects are not expected to be significant for the next generation of detectors but it is nevertheless useful to be able to quantify them. Our results show that 
(perhaps surprisingly) internal stratification has no impact whatsoever on the Love number. We also show that crust elasticity provides a (predictably) small correction to existing models. 
\end{abstract}

\maketitle

\section{Introduction}\label{sec:intro}

When a massive body is exposed to a relatively weak external gravitational field we expect it to respond by changing shape. This is most easily understood by considering the gravitational effects of the Moon on the Earth. The oceans move to reach equilibrium as the moon orbits, leading to the observed tides. The effect will also deform the Earth's elastic crust, again to reach an equilibrium with the passing body. The deformation of the massive body modifies the tidal potential, an effect that can be expressed in terms of dimensionless quantities known as the Love numbers. We will consider this problem for a neutron star in a close binary system, focussing on the tidal Love number; the measure of the tidal response of an object to an external gravitational field \cite{Stacey}. 

Specifically, we are interested in $k_2$, the measure of a tidal deformation by a quadrupole perturbation due to an external gravitational field. This quantity is of direct relevance for future gravitational-wave astronomy, as it provides a potentially detectable ``finite-size correction'' to the late stages of a binary inspiral signal \cite{Flanagan:2008}.  The effect depends on the density distribution of the star, and hence may be used as a diagnostic to use observational data to constrain neutron star theory. The implications for observations, and the formalism required to work out the quadrupole deformation in relativity, were first discussed in  \cite{Flanagan:2008}. Since then the formalism has been generalized to determine, in particular, the shape Love number, see for example \cite{Poisson:2009,Damour:2009}. Applications have mainly considered barotropic fluid models, with realistic equations of state studied in \cite{Damour:2009} and the difference 
between neutron stars and self-bound quark stars being quantified in~\cite{Postnikov:2010}. 

The main aim of the present work is to develop the formalism required to study the tidal deformation of real neutron stars, accounting for key aspects like the elastic crust and variations in the interior composition. These are important developments, even though we do not expect these features to be easily observed in a binary gravitational-wave signal. The Love number leaves an imprint that is borderline detectable by advanced LIGO \cite{Flanagan:2008}. The effect will be more important for the third generation Einstein Telescope \cite{READ}, but small corrections to it may not seem that relevant. Nevertheless, as a matter of principle it is important to  move beyond back-of-the-envelope estimates and quantify the actual effect. Moreover, developments in this direction are required to address a number of questions of more immediate astrophysical relevance. By developing the required relativistic perturbation formalism for the static deformations induced by the tidal interaction, we prepare the ground for studies of general crust asymmetries, e.g., neutron star ``mountains'' which lead to gravitational-wave emission at twice the star's spin frequency \cite{Jaranowski:1998,LIGO}. The work presented here represents key steps towards modelling such mountains in general relativity, which would allow us to consider realistic equations of state for the first time in this context. A closely related problem to that considered here concerns the breaking of the crust during binary inspiral. It is easy to estimate (based on the energetics involved~\cite{Postnikov:2010}) that the crust will break before the final merger, but when exactly does this happen and where does the crust yield first? Another related problem concerns the quasiperiodic oscillations observed in the tails of magnetar flares \cite{Israel,Watts:2007}. Again, a key issue concerns the build-up of stresses in the crust of these strongly magnetized stars, the eventual fracture and the induced dynamics. The formalism developed here provides a useful first step for a discussion of this problem as well.

\section{Neutron star relasticity}

The structure of the neutron star crust is relatively well understood  \cite{LivRevCrust}, although issues associated with the dynamics of the 
superfluid neutron component and the possible pasta phases in the deep crust layers need to be explored further. The theory required to model 
the crust in the framework of general relativity is also well developed. Building on pioneering work by Carter \cite{Carter:1972,Carter:1973} and more recent efforts by  Karlovini and Samuelsson~\cite{Karlovini:2003}, two of us have recently completed a formalism for Lagrangian perturbations of a relativistic elastic system \cite{RELLAG} (see  \cite{Andersson:2011} for the corresponding model in Newtonian gravity).  These developments allow us to consider a range of relevant astrophysical applications. 

In order to quantify the effect that the neutron star crust has on the tidal deformability of the star, we need to formulate and solve the problem for static (quadrupole) deformations of a given stellar model. As usual, this is a two-stage process. The first stage is straightforward. If we assume that the unperturbed elastic star is relaxed, i.e., that the crust is unstrained, then we simply need to solve the usual Tolman-Oppenheimer-Volkoff (TOV) equations for a non-rotating relativistic perfect fluid star. The crust manifests itself only at the linear perturbation level.  At this second stage, we need to consider the static perturbations of a model that accounts for the associated elastic stresses. 

In this section, we develop the formalism required to solve the tidal deformation problem for elastic relativistic stars. In particular, we develop the relevant perturbation equations and discuss the key differences from the perfect fluid case. Throughout most of the analysis, we use geometric units where $c=G=1$, and all primes denote differentiation with respect to the radial coordinate. We restore the physical units only when they are useful for analysis purposes.

\subsection{The background problem}\label{sec:background}

The equilibrium of a static, spherically symmetric, relativistic star is described by a spacetime metric $g_{ab}$ given by 
\be
ds^2 = -e^{\nu(r)} dt^2 + e^{\lambda(r)} dr^2 + r^2 d \theta^2
    + r^2 \sin^2 \theta d \varphi^2 \  .
\ee
The fluid four-velocity is simply given by
\be
u^a = e^{- \nu/2} t^a \ ,
\ee
where $t^a=(\partial_t)^a$  is the timelike Killing vector of the spacetime. 

We want to model a neutron star with a fluid core, an elastic crust and possibly a fluid ocean. The corresponding equilibrium problem simplifies significantly if we assume that the crust is relaxed in the unperturbed configuration. If this is the case, then the problem reduces to that of a fluid system.
Physically, this assumption makes sense provided the crust has had time to release any built up strain, e.g. via plastic flow,
before the tidal interaction with the star's companion becomes sizeable. The upshot is that we can use the perfect fluid stress-energy tensor to model the background star.
In other words, we have
\be
T_{ab} = (\ep+ P) u_a u_b + P g_{ab} = \ep u_a u_b + P\perp_{ab}\ ,\label{Tfluid}
\ee
where $\perp_{ab} = g_{ab}+u_au_b$ is the usual projection orthogonal to the fluid frame.
In the particular case of a barotropic star, the pressure and energy density are related by an equation of state of form $P = P(\ep)$.
We will discuss the more general case, where the (cold) equation of state is such that pressure also depends on the composition of matter at supranuclear densities, later.

Once the equation of state is provided we need to solve the standard TOV equations starting with
\be
\lambda' = - \frac{2 e^\lambda}{r^2} \left( M - 4 \pi r^3 \rho \right)  \ ,
\label{lap}\ee
where the gravitational mass $M(r)$, inside radius $r$, is obtained from
\be
M^\prime = 4\pi r^2\ep \ . \label{eq:dMdr}
\ee
Then using
\be
\nu' = \frac{2 e^\lambda}{ r^2} \left( M + 4 \pi r^3 P \right) = - \frac{2}{(\ep+P)} P' \ ,
\label{GR_sph:dnudr}
\ee
we arrive at
\be
P^\prime = - \frac{(\ep+P)(M+4\pi r^3P)}{r(r-2M)} \ .
\label{eq:dPdr}
\ee
These relations provide enough information to determine the background stellar model.

\subsection{Static perturbations of a fluid star}\label{sec:static_perfect}

In order to investigate the tidal deformations, we need to consider static perturbations of the system. 
Since we plan to account for elastic contributions, it is natural to approach the problem within Lagrangian perturbation theory. 
The relevant formalism has recently been developed \cite{RELLAG}. For the present purposes it is worth noting that the problem we consider is
 very similar to that of the zero-frequency subspace of perturbations for slowly rotating stars. This means that we can draw on work relating to the gravitational-wave driven r-mode instability. In particular, the spacetime part of the perturbation equations is identical to that considered by \cite{Lockitch:2003}.

Due to the nature of the tidal interaction, we focus on the polar perturbations. These lead to the electric-type Love numbers, as discussed in \cite{Poisson:2009,Damour:2009}. We use the standard Regge--Wheeler gauge, representing the perturbed metric $h_{a b}$ by
\be
h_{a b} = \begin{pmatrix}
H_0(r) e^\nu & H_1(r) & 0 & 0 \\ 
\mathrm{symm} & H_2(r) e^\lambda & 0 & 0 \\
0 & 0 & r^2 K(r) & 0 \\
0 & 0 & 0 & r^2 \sin^2 \theta K(r) 
          \end{pmatrix} Y_l^m \ ,
 \ee
(noting a typo in the off-diagonal element in \cite{Lockitch:2003}).
Due to the spherical symmetry of the problem, the different multipoles do not couple. In fact, we can, without loss of generality, assume that the perturbations are axisymmetric (set $m=0$).

We introduce a static displacement vector given by \footnote{We note for clarity that this is different from the displacement vector used by \cite{Thorne:1967} and subsequent works such as \cite{Finn:1990}.},
\be
\xi^a =\left[ {\frac{1}{r}} W(r) Y_l^m r^a + V(r) \nabla^a Y_l^m \right] \ .
\ee
Within relativistic Lagrangian perturbation theory \cite{Carter:1973,FS75,Friedman:1978a,LivingRevNils,RELLAG} we then have the perturbed four-velocity
\be
\Delta u^a = {\half} u^a u^b u^c \Delta g_{b c} \ , 
\ee
where
\be
 \Delta g_{a b}  = h_{a b} + 2 \nabla_{(a} \xi_{b)} \ .
\ee
Moreover,  the Lagrangian perturbations are related to the Eulerian perturbations by 
\be
 \Delta = \delta + \mathcal{L}_\xi \ ,
\ee
where $\mathcal{L}_\xi$ is the Lie derivative along $\xi^a$. 
It follows that the perturbed 4-velocity, $\delta u^a$, is given by
\be
\delta u^a = \perp^a_{\ \ b} \mathcal{L}_u \xi^b + {\half}  u^a u^b u^c h_{b c} \ .
\ee
Since the displacement vector is taken to be static, we have
\be
\delta u^t = {\half} e^{-\nu/2} H_0 Y_l^m \ , \qquad \mbox{and} \qquad \delta u^j = 0 \ ,
\ee
where $j=1-3$ represents a spatial index.

We also know that the perturbed number density can be obtained from the displacement vector,
since~\cite{LivingRevNils}
\beq
\Delta n = \delta n + \xi^a \nabla_a n =  - \half n \perp^{ab} \Delta g_{ab} \ . \label{eq:deln} 
\eeq
Explicitly, we have
\begin{align}
\Delta n = - \frac{n}{r^2} &\left[ r^2 \left( K + \half H_2 \right) - l(l+1) V +  rW' + \left(1+ \half r \lambda'\right)W  \right] Y_l^m  \ . \label{eq:deln2}
\end{align}
In the case of barotropic matter the energy density will be a function only of the number density of the constituent particles, $\rho=\rho(n)$ which means that, 
\be
\Delta \rho = {\frac{d \rho}{dn}} \Delta n = \mu \Delta n \ , 
\ee
where we have introduced the chemical potential $\mu$ \cite{LivingRevNils}. 
For algebraic simplicity later we define
\begin{align}
\perp_g = \frac{2}{r^2} &\left[ r^2 \left( K + \half H_2 \right) - l(l+1) V  +  rW' + \left(1+ \half r \lambda'\right)W  \right] \ .\label{perpg}
\end{align}
Combining \eqref{eq:deln2} with the Gibbs relation $P + \rho = n \mu$, we see that 
\be
\Delta \rho = - \half \left( P + \rho \right)\perp_g Y_l^m \ .
\ee
Finally, in the case of a barotrope, the perturbed pressure follows from
\be
\Delta P = {\frac{dP}{d\rho}} \Delta \rho = c_s^2  \Delta \rho \ , \quad  \text{and} \quad \delta P = {\frac{dP}{d\rho}} \delta \rho\ ,\label{eq:drhoP}
\ee
where we have defined the speed of sound, $c_s^2$.
These relations show that we can choose to work either with the components of the displacement vector or the perturbed density/pressure.
One of the variables, $\delta \rho$, $W$ and $V$ is redundant. 
The situation is, of course, different for a non-barotropic model. We will discuss this case later.

To complete the specification of the fluid problem we need the perturbed Einstein equations (for $l\ge 2$). The right-hand side of the equations
follows from \eqref{Tfluid}. In the case of a perfect fluid we have
\begin{align}
\delta T_a^{\ b} = \left( \delta \ep + \delta P \right) &u_a u ^b + \delta P \delta_{a}^{\ b} + (P + \ep) \left( u^b \delta u_a +  u_a \delta u^b\right) \ , 
\label{pT1}
\end{align}
where the scalar perturbations are to be expanded in spherical harmonics. It is worth noting that all off-diagonal components vanish identically since $\delta u^j =0$.

The corresponding left-hand side of the Einstein equations, expressed in terms of the perturbed metric, obviously
does not depend on the matter content of the spacetime at all. Basically, the required expressions 
can be found in a number of other studies, including \cite{Lockitch:2003,Thorne:1967} and \cite{Hinderer:2008}. In the case of a spherically symmetric 
system, we end up with six coupled ordinary differential equations to solve. The final fluid equations are easily obtained from the full elastic equations 
discussed below, so we will not give them here.

\subsection{Elasticity}\label{sec:elasticity}

Having outlined the fluid perturbation problem, and the foundations of Lagrangian perturbation theory in the relativistic setting, we can move on to consider the role of the neutron star's elastic crust. As we are working with the Einstein equations, we only need to consider the relevant alterations to the stress-energy tensor. A perfect fluid cannot support shear stresses and hence does not introduce off-diagonal stress-energy terms. 
When we account for elasticity we need to include such stresses \cite{RELLAG, Andersson:2011}. 

As discussed above, we assume the background star is in a relaxed unstrained state. Given this assumption, the elastic crust does not contribute to the stress-energy tensor of the background star, which leaves our equilibrium equations unaltered. The elastic contributions enter only through the perturbed stress-energy tensor.  Following \cite{RELLAG}, we have the Lagrangian perturbation of the anistropic stress tensor;
\be
\Delta \pi_{a b}= - 2 \muc \Delta s_{a b} \ ,
\ee
where
$\muc$ is the shear modulus (not to be confused with the chemical potential), and 
\be
 2 \Delta s_{a b}  = \left(  \perp^c_{\ a} \perp^d_{\ b}- \frac13 \perp_{ab} \perp^{cd} \right) \Delta g_{c d}  \ .
\ee
In the case of an unstrained background, 
these expressions lead to
\be
\delta \pi_a^{\ b} = -2 \muc \left( \perp^c _{\ a} \perp^{d b} - \frac13 \perp_{a}^{\ b} \perp^{cd }  \right)\Delta g_{cd} \  ,
\label{dpab}\ee
which should be added to the fluid result \eqref{pT1}.
For future reference, it is useful to note that
\be
\delta \pi_t^{\ b} = 0 \ .
\ee

In order to complete the perturbed Einstein equations in the elastic case, we need to express the perturbed stress tensor in terms of our chosen 
variables. Thus, we find that 
\begin{align}
\delta \pi_r^{\ r} &\rightarrow {\frac43 r^2} \muc \left[ r^2(K-H_2) - l(l+1)V - 2r W' +\left(4-r\lambda' \right)W \right] \ , 
\end{align}
(here, and in the following, we suppress the angular dependence for clarity).
Moreover, since the anisotropic stress must be trace-free, c.f.,  \eqref{dpab}, it follows immediately that
\be
\delta \pi_\theta^{\ \theta}+\delta \pi_\varphi^{\ \varphi} = -\delta \pi_r^{\ r} \ .
\ee
Note that this result requires that the background is relaxed.
That is, we have
\begin{align}
\delta \pi_\theta^{\ \theta}&+\delta \pi_\varphi^{\ \varphi} \rightarrow \frac43 \muc \left[ r^2(H_2-K)+l(l+1)V+2rW'- \left( 4-r\lambda'\right)W \right] \ .
\end{align}
We also find that
\be
\delta \pi_\theta^{\ \theta}-\delta \pi_\varphi^{\ \varphi} \rightarrow 8 \muc V \ ,
\ee
and, finally,
\be
\delta \pi_r^{\ \theta} \rightarrow - { \frac{2\muc}{r^3}} \left( r V' - 2V + e^\lambda W\right) \ .
\ee

Given the above results, the perturbation equations for the stellar interior become more complicated. However, the $[t\;t]$-component of the perturbed Einstein equations is unaffected by the presence of the crust, so we have from the fluid case,
\be
 e^{-\lambda} r^2 K''
+ e^{-\lambda} \left(3- \frac{ r\lambda'}{ 2} \right) rK' - \left[ \half l(l+1)-1\right] K  
 - e^{-\lambda} rH_2'
-\left[\half l(l+1)+ e^{-\lambda} \left( 1 - r \lambda'\right) \right]H_2
 = - 8\pi r^2\delta\ep \ .
\label{tt}
\ee
Moreover, because our star is spherically symmetric and $\delta \pi_t^{\ r} =0$, we find  that $H_1=0$ (c.f., \cite{Hinderer:2008}). Finally, because $\delta \pi_a^{\ b}$ is traceless, we can take the trace of the perturbation equations to arrive at a second equation without explicit dependence on the elasticity. This equation takes the form
\begin{multline}
-r^2 H_0'' +\left( \frac12 r \lambda' - r\nu'-2\right)r H_0' + r^2 \nu' K' -\frac12 r^2 \nu' H_2' + l(l+1)e^\lambda H_0\\
+\left[ 2\left( e^\lambda-1 \right) - r \left(\lambda'+3\nu'\right)\right]H_2
= 8\pi r^2 e^\lambda(3\delta P + \delta \rho) \ .
\label{eq:track}
\end{multline}

In practice, it is useful to introduce two additional variables closely related to the traction. These variables, which vanish in the fluid case, are key to the 
implementation of the relevant junction conditions at the crust-core interface. Hence, we define a variable linked to the radial component of the traction;
\be
T_1 \equiv  \frac43 \muc \left[ r^2(K-H_2) - l(l+1)V - 2r W' +\left(4-r\lambda' \right)W \right]  \ , 
\label{T1def}
\ee
and its $\theta$ component;
\be
T_2 \equiv -4 \muc \left( r V' - 2V + e^\lambda W\right)\label{T2def}\ .
\ee

Working out the perturbed Einstein equations, 
following the same strategy as in the fluid problem (see  \cite{Lockitch:2003} and \cite{Hinderer:2008}), we find that the difference between the $[ \theta\; \theta]$ and the $[ \varphi\; \varphi]$ components 
 leads to
\be
H_2 - H_0 = 64\pi \muc V \ .
\label{elH2}\ee
Meanwhile, the $[r\;\theta]$ equation leads to 
\be
 K' =   e^{-\nu} \left[e^{\nu} H_0 \right]' + {\frac{32 \pi \muc}{r}} \left(  r\nu' + 2 \right) V  - 8 \pi \frac{T_2}{r} \ .
\label{eldK}\ee
In terms of the traction variables we can write the   sum of the $[ \theta\; \theta]$ and the $[ \varphi\; \varphi]$  
 component as
\be
 \delta P = {\frac{1}{16 \pi r}} e^{-\lambda} \left( \nu' + \lambda' \right) H_0  
  + { \frac{e^{-\lambda}}{r^2}}
\left\{\{ 2\muc e^\lambda \left[ 2 - l(l+1) \right] V - \left[ {\frac{r}{4}} \left( \nu' - \lambda' \right) + 1 \right] T_2 
+ \frac{e^\lambda}{2} T_1 - \frac{T_2'}{2} + \frac{T_2}{2}  \right\} \ ,
\label{eldp}
\ee
where we have also used \eqref{elH2} and \eqref{eldK}.

Finally, the $[r\;r]$ component of the Einstein equations leads to 
\begin{multline}
\left[ l(l+1)-2\right] e^\lambda K = r^2 \nu' H_0' +  \left[ l(l+1)e^\lambda - 2 + r^2 (\nu')^2 - r \left( \nu' + \lambda' \right) \right] H_0 \\
+ 32\pi \muc \left[ r^2 \left( \nu'\right)^2 + l(l+1) - 2 \right] V - 4 \pi r \left( \nu' + \lambda'\right) T_2 + 8 \pi ( rT_2'-T_2) - 24 \pi T_1 \ .
\label{elK}
\end{multline}

It is worth noting that, 
by adding \eqref{eldp} and \eqref{elK} then subtracting \eqref{tt} we have the trace of the Einstein equations, i.e., we recover \eqref{eq:track}.

\subsection{The final perturbation equations}\label{sec:crust_eqns}

The set of equations that we need to solve in the crust region combines the conservation law \eqref{eq:deln2} with the perturbed Einstein equations \eqref{elH2}--\eqref{elK}, including the definitions \eqref{T1def} and \eqref{T2def}. We want to formulate the problem in such a way that the required integration becomes as straightforward as possible. To do this, we note that we can reduce the order of the system. However, in the elastic case this reduction does not lead to the same level of simplification as in the fluid problem, where we only need to solve a single second order
equation for $H_0$, c.f., \eqref{eq:master} below. This is of course as expected, as the problem now has additional degrees of freedom due to the elasticity. For example, it is natural that we will have to solve for the components of the displacement vector. We take the view that these components are obtained from the definitions \eqref{T1def} and \eqref{T2def}. That is, we integrate 
\begin{align}
W' - \left( {\frac{2}{r}} - {\frac{\lambda'}{2}} \right) W &= \frac{r}{2} (K-H_0) - \left[ 32\pi \muc r + \frac{l(l+1)}{2r} \right] V - \frac{3}{ 8 \muc r} T_1 \ , 
\label{deq1}
\end{align}
and
\be
V' - {\frac{2V}{r}} = -{\frac{e^\lambda}{r}} W - {\frac{1}{4 \muc}} \frac{T_2}{r} \ .
\label{deq2}
\ee

Next we note that we have several relations involving the perturbed pressure. If we focus on the barotropic problem it follows from 
\eqref{eq:deln2} that we should have
\be
\Delta P = c_s^2 \Delta \rho = - \half (P+\rho) c_s^2 \perp_g \ , 
\ee  
where
\be
\Delta P = \delta P + \frac{W}{r} P' = \delta P - \frac{1}{2 r} (P+\rho) \nu' W \ .
\ee
Using \eqref{deq1} to remove $W'$ from $\perp_g$ we arrive at the algebraic relation
\begin{align}
\delta P = -\frac{P+\rho}{r^2} c_s^2 &\left\{ \frac32 r^2  K - \frac{3}{2} l(l+1)V + \left( 3 - \frac{r \nu'}{2 c_s^2} \right)  W
- \frac{3}{8 \muc } T_1 \right\}  \label{alg1} \ .
\end{align}
We can rearrange this as an equation that determines $T_1$, giving
\begin{align}
T_1 = \frac{8 \muc}{3} &\left\{ \frac{r^2}{(P+\rho) c_s^2 } \delta P + \frac32 r^2  K - \frac{3}{2} l(l+1)V + \left( 3 - \frac{r \nu'}{2 c_s^2} \right) 
W \right\} \ .
\label{alg2}
\end{align}

By combining \eqref{eldp} and \eqref{elK}, in such a way that $T_2'$ is removed, we arrive at a second algebraic relation. This provides another
expression for $\delta P$, giving
\begin{align}
16 \pi r^2 e^\lambda \delta P &= r^2 \nu' H_0' + [2-l(l+1)]e^\lambda K + \left[ l(l+1)e^\lambda - 2 +r^2 \left(\nu'\right)^2 \right] H_0 \nonumber\\
&+ 32 \pi \muc \left\{ r^2 \left( \nu'\right)^2 + [l(l+1)-2]\left( 1 - e^\lambda \right) \right\} V - 8 \pi \left( r\nu' + 2 \right) T_2 + 8\pi \left( e^\lambda -3 \right) T_1 \ .
\label{alg3}
\end{align}
We can use this for the left-hand side of \eqref{eldp}, which then becomes an equation that we can integrate
to get $T_2$;
\begin{align}
T_2' + \left[ \frac12 \left( \nu'-\lambda'\right) + \frac{1}{r} \right]T_2 &= -2e^\lambda r \delta P + \frac{1}{8 \pi} \left( \nu'+\lambda'\right) H_0 + \frac{4\muc}{ r } e^\lambda \left[ 2 - l(l+1) \right] V + \frac{e^\lambda}{ r } T_1 \ .
\label{deq3}
\end{align}

Finally, we have \eqref{eldK} for $K'$, which we can use in \eqref{eq:track} (together with \eqref{elH2} for $H_2$) to get
\begin{multline}
-r^2 H_0'' +\left[ \frac12 r (\lambda' - \nu')-2\right]r H_0' + \left\{ l(l+1)e^\lambda + 2(e^\lambda-1) -r(\lambda'+3\nu') +r^2 \left(\nu'\right)^2 \right\} H_0 \\
= 8\pi \left\{ r^2 e^\lambda(3\delta P + \delta \rho) + 16 \muc \left[ 1 - e^\lambda +r\left( \nu' + \frac12 \lambda'\right) -\frac14 \left(r\nu'\right)^2\right]V +4r^2 \nu' \left( \muc V\right)' +r\nu' T_2
\right\} \ .
\label{deq5}
\end{multline}
These equations completely specify the perturbation problem in the elastic region. It is easy to verify that we have the same number of equations as we have unknowns.

\subsection{Interface conditions}\label{sec:BC}

Having determined the perturbation equations for the elastic crust region, we need to connect them to the fluid perturbations for both the core and the fluid ocean near the star's surface. This requires a set of interface conditions at a radius $r_c$, which could represent either the crust-core or the crust-ocean interface. The interface problem has already been analyzed in detail, the most relevant work being that of Finn \cite{Finn:1990}. As our analysis is identical, we simply summarize the results here. 

From our assumption that the unperturbed crust is in a relaxed state we know that all background quantities are continuous across a fluid-crust interface. This includes the density. It may, of course, be that the true equation of state is such that the interface is associated with a small density discontinuity (c.f., the discussion in \cite{Postnikov:2010}). We do not account for this possibility here, but it would be very easy to do so, should it be required. 

To determine the behaviour of the perturbed quantities across the fluid-elastic interface we calculate the intrinsic curvature at the interface. From the Israel junction conditions we know that the intrinsic curvature must be continuous. In the fluid problem (in fact, even for multi-fluids \cite{Andersson:2002}),  this results in the continuity of all perturbed metric variables, as well as their first derivatives. The proof of this relies on the fact that  $H_2=H_0$ in the fluid case.
This is, however, no longer true when we consider the elastic problem.   Consequently, the behaviour at a fluid-elastic interface is a little bit more complicated.

Nevertheless, we know from the analysis of Finn  \cite{Finn:1990} that the continuity of the first fundamental form demands the continuity of 
$W$, $H_0$, $H_1$ and $K$ at the interface. The first of these conditions follows from the continuity of $\xi^r=W/r$, which tells us that in order to ``avoid a void'', the radial displacement component $W$ must be continuous across the interface. Note also that the condition on $H_1$ is irrelevant to us as this variable vanishes identically in the problem under consideration.

Combining \eqref{elH2} with the stated continuity conditions we see that we should have
\be
 \left[H_2\right]_{r_c}= 32\pi\left[\muc V\right]_{r_c}\ ,\label{eq:conH2}
\ee
where $[A]_{r_c}$ is shorthand for $\lim_{\epsilon\rightarrow0}A(r_c+\epsilon)-\lim_{\epsilon\rightarrow0}A(r_c-\epsilon)$.
This result shows that we should expect a jump in the $H_2$ perturbation; first of all the shear modulus will vanish sharply as the 
crust gives way to the fluid core, and secondly there is no reason why the tangential displacement $V$ should be continuous (at least not as long as we are ignoring viscosity and magnetic field stresses). 
In the limit where $\muc\rightarrow0$ we obviously recover the standard fluid interface condition.

The remaining interface conditions can be obtained either by imposing the continuity of the extrinsic curvature $K_{ab}$, or 
the surface stresses $S_{ab}$ on the perturbed hypersurface. The two approaches are related since \cite{PoissonAGR},
\be
S_{ab}= \frac{1}{8\pi}\left([K_{ab}]-[K]\preup{(3)}{g}_{ab}\right) \ ,
\ee
where $\preup{(3)}{g}_{ab} = g_{ab}-N_a N_b$ is the induced three-metric on the surface and $K = K_{ab}\preup{(3)}{g}_{ab}$, and $N_a$ is the normal to the surface. Since the right-hand side must be continuous across the surface we demand that $[S_{ab}]=0$. This provides two additional conditions \cite{Finn:1990};
\be
 \left[ T_1 + r^2\Delta P \right]_{r_c}=0\ ,\label{eq:conTr}
 \ee
 for the  (perturbed) radial traction, and 
 \be
 \left[T_2\right]_{r_c}=0\ ,\label{eq:conT2}
\ee
for the horizontal part. 
Since the background variables are taken to be continuous across the interface, and since the intrinsic curvature conditions lead to $[W]_{r_c}=0$, condition \eqref{eq:conTr} yields
\be
 \left[\delta P + T_1\right]_{r_c}=0.\label{eq:conT1}
\ee

We now have six conditions at each fluid-crust interface, to combine with the six ODEs in the crust. The problem is therefore well posed.

The implementation of the interface condition depends, to some extent, on the physical model considered. We are interested in a stellar model with a fluid core and an elastic crust that transitions back to a fluid ocean near the surface. Consequently, we need to impose the junction conditions at two fluid-elastic interfaces. Integrating from the centre of the star, we solve the fluid problem up to the crust-core interface. At this point the conditions provide us with 
all information needed to continue the integration using the elastic equations. Finally, the same conditions are imposed at the crust melting point. 
Here, the continuity of $H_0$ and $K$ implies that we have the information required to integrate the fluid equations to the actual surface of the star.  

Specifically, we first of all have the continuity of  $H_0$ and $K$ at each interface. Moreover, from the condition $[T_2]=0$ we realize that since the shear modulus $\muc$ has a finite value in the crust, but vanishes in the fluid, we must have $T_2=0$ at the interface. The condition $[W]=0$ is a little more complicated.
Since we do not calculate the radial perturbation, $W$, in the fluid we need a means to initiate the integration in the crust.
To do this, we make use of the condition \eqref{eq:conTr}.
Considering the algebraic relations between $\delta P$ and $T_1$, Eqns. \eqref{alg1} and \eqref{alg2}, and using continuity we have
\begin{align}
16\pi r^2e^{\lambda}\delta P_E = r^2\nu^\prime H_E^\prime &+ \left[2-l(l+1)\right]e^{\lambda}K_E + \left[l(l+1)e^\lambda - 2 + r^2(\nu^\prime)^2\right]H_E \nonumber\\
&+ 32\pi\muc\left\{r^2(\nu^\prime)^2+[l(l+1)-2](1-e^\lambda)\right\}V_E+ 8\pi(e^\lambda-3)T_{1E}   \ , 
\label{eq:Wc1}
\end{align}
where the subscript $E$ denotes a quantity calculated in the crust region. Using the jump conditions we can express $H_E$ and $K_E$ in terms of the fluid quantities.
We also use
\be
 \frac{3T_{1E}}{8\muc} = \frac{r^2\delta P_E}{(\rho+P)c_s^2} + \frac{3}{2}r^2K_E - \frac{3}{2}l(l+1)V_E + \left(3-\frac{r\nu^\prime}{2c_s^2}\right)W_E\label{eq:Wc2} \ .
\ee
The continuity of the radial traction closes this set of equations: since $T_{1F}=0$ ($F$ indicating a variable in the fluid region) identically we must have 
\be
 T_{1E} = r^2\left[\frac{1}{2}(\rho+P)H_F - \delta P_E\right]\label{eq:Wc3}\ .
\ee
Combining equations \eqref{eq:Wc1}--\eqref{eq:Wc3} we have the information required to determine $W_E$, the radial perturbation at  the base of the crust.

\section{The Love Number} \label{sec:Love}

In order to implement the formalism in an astrophysically meaningful context, we will quantify how the crust elasticity affects the 
tidal deformation of a neutron star. This effect can be expressed in terms of the 
tidal Love number. A static spherically symmetric star of mass $M$ and radius $R$ exposed to a time-independent external tidal field $\mathcal{E}_{ij}$ will develop a quadrupole moment $Q_{ij}$. To linear order we relate this quadrupole moment $Q_{ij}$ to the tidal moment $\mathcal{E}_{ij}$ thus defining the Love number, $k_2$, \cite{Hinderer:2008},
\be
 Q_{ij} = \frac23k_2R^5\mathcal{E}_{ij} \ .\label{eq:Love}
\ee

We briefly review the procedure used to calculate the tidal Love number below. For a detailed description we refer the reader to any of \cite{Flanagan:2008,Poisson:2009,Damour:2009} or \cite{Hinderer:2008}.

Following \cite{Hinderer:2008}, the Love number is extracted from the asymptotic behaviour of the gravitational field of a tidally deformed body. To do this we note that the vacuum perturbation problem reduces to a single ODE for $H_0$
\be
H_0^{\prime\prime}+\left( \frac{2}{r} - \lambda^\prime \right) H_0^\prime - \left[ \frac{l(l+1)e^\lambda}{r^2} - (\lambda^\prime)^2 \right] H_0 = 0 \ ,
\ee
where we have used $\nu = -\lambda$ and $M(r) = M=\rm{const.}$ exterior to the star. This equation may be solved in terms of the associated Legendre polynomials \cite{Flanagan:2008,Damour:2009,Thorne:1967}, leading to;
\be
H(r) = a_P P_{l2}\left(\frac{r}{M}-1\right)+a_Q Q_{l2}\left(\frac{r}{M}-1\right).\label{eq:Htop}
\ee
We have recorded both the decreasing solution $P_{lm}(x)$ and growing solution $Q_{lm}(x)$ for $m=2$.

Since the problem is studied within linear perturbation theory the amplitude of the solution is arbitrary, so we 
demand that the function 
\be
y(r) = r\frac{H^\prime(r)}{H(r)},\label{eq:y}
\ee
matches across the surface at the star. Substituting the general solution Eqn.~\eqref{eq:Htop} into Eqn.~\eqref{eq:y}, we get
\be
 y(x)=(1+x)\frac{P_{l2}^\prime(x)+a_lQ_{l2}^\prime(x)}{P_{l2}(x)+a_lQ_{l2}(x)}\ ,\label{eq:Damour}
\ee
where $x=R/M-1$.
Following, \cite{Damour:2009} we have defined $a_l=a_Q/a_P$ which is determined by matching Eqn.~\eqref{eq:Damour} across the surface of the star. This leads to 
\be
 a_l = -\left.\frac{P_{l2}^\prime(x)-Cy_lP_{l2}(x)}{Q_{l2}^\prime(x)-Cy_lQ_{l2}(x)}\right|_R\ ,\label{eq:Damourcoef}
\ee
where $C=M/R$ is the compactness of the star. Substituting $l=2$, and taking the asymptotic expansion we get
\be
H_0 \approx \frac85 \left( \frac{M}{r} \right)^3 a_P + 3 \left( \frac{r}{M} \right)^2 a_Q \ , 
\ee
which is the result used in \cite{Hinderer:2008}.

Finally, we relate the coefficient $a_l$ to the Love number by comparing to the response of a spherically symmetric star to an external quadrupolar field, $\mathcal{E}_{ij}$ \cite{Hinderer:2008, Thorne:1998},
\be
 \frac{1-g_{tt}}{2} = -\frac{M}{r} - \frac{3Q_{ij}}{2r^3}\left(\frac{x^i x^j}{r^2}-\frac{1}{3}\delta^{ij}\right) + \frac{1}{2}\mathcal{E}_{ij}x^i x^j \ , 
\ee
where we have dropped terms of order $O(1/r^3)$ and $O(r^3)$. Using 
\be
 g_{tt} = -\left(1-\frac{2M}{r}\right)\left( 1 - H_0 Y_{lm} \right)\ , 
\ee
and Eq.~\eqref{eq:Love}, we determine the Love number, $k_2$, to be
\be
k_2 = \frac{4G}{15} \left( \frac{M}{R}\right)^5 a_2 \ ,
\ee
where $a_2$ is determined by Eqn.~\eqref{eq:Damourcoef} from the solution to the interior problem.

In the perfect fluid problem, i.e., ignoring the elastic contributions, the interior equations can be reduced to a single second order equation for $H_0$ \cite{Hinderer:2008},
\be
r^2 H_0^{\prime\prime}+\left[ \frac{2}{r} +  \half \left(\nu^\prime - \lambda^\prime \right)\right] r^2 H_0^{\prime} 
+  \left[ 2\left( 1 - e^\lambda\right)  - l(l+1) e^\lambda +2 r (2\nu' + \lambda') - r^2 \left( \nu'\right)^2 \right]H_0
= -8\pi r^2  e^\lambda(\delta P + \delta \rho)\label{eq:master}\ .
\ee
Combining this equation with $\delta P = c_s^2\delta \rho$ (for a barotropic model) and \eqref{eldp}, obviously still ignoring the elastic contributions, we have a complete formulation of the problem, and a route to determining the coefficients $a_P$ and $a_Q$ required in the exterior. When considering the elastic system, we have to solve the set of equations discussed in the previous section in order to determine $a_P$ and $a_Q$.

For future reference, it is worth noting that \eqref{eq:master} contains the same information as \eqref{eq:track}, as is necessary for the problem not to be
overdetermined. However, one can show that this implies that 
\be
8 \pi r^3 e^\lambda \nu' \delta \rho = \left[ r^2 \lambda'' - r^2 \left( \lambda'\right)^2 + 2\left(1-e^\lambda \right)  \right] H_0 \ .  
\ee
Making use of the background equations and the sum of the $[ \theta\; \theta]$ and  $[ \varphi\; \varphi]$   components of the Einstein equations we find that we must have
\be
\rho^{\prime} \delta P = P^{\prime} \delta \rho  \ .
\label{prhorel}\ee
This is (obviously) true for barotropes, where $P = P(\rho)$, so the problem is well posed.

Finally, 
using Eqn.~\eqref{eq:y} and the compactness parameter $C$, we determine the coefficient $a_2$ and obtain the Love number from
\begin{multline}
k_2 = \left\{\frac{8C^5}{5}(1-2C)^2[2+2C(y-1)-y]\right\}\times\\
  \left\{2C(6-3y+3C(5y-8))+4C^3[13-11y+C(3y-2)+2C^2(1+y)]\right.\\
\left.+3(1-2C)^2[2-y+2C(y-1)]\log{(1-2C)}\right\}^{-1} \ . 
\end{multline}

\section{Results}

The previous sections specify the calculations needed to account for the crust elasticity, determine tidal response 
and quantify the Love number $k_2$ for realistic neutron star models. We will now show how the detailed interior physics affects the result. 
We consider both
composition variations and the role of the elastic crust.

\subsection{Stratification}\label{sec:stratification}

It is well-known that variations in the composition of the neutron star core, e.g. represented by a varying proton fraction, may have repercussions for the global dynamics. In particular, the associated stratification may lead to the presence of g-modes in the star's oscillation spectrum \cite{RG92}. From a mathematical point of view, these effects arise from the equation of state depending on two (or more) parameters. The impact on the dynamics depends on 
the detailed timescales of the problem. We will consider two extreme limits. In the first case,  when reactions are very fast, 
the perturbed fluid elements will adjust to their surroundings very quickly. In other words, they lose their original identity as the system evolves.
In this limit one would expect the problem to remain effectively barotropic.
In the opposite limit, 
reactions act slowly compared to the dynamics. 
In the present context, this should be taken as meaning that the reactions are slow compared to the binary inspiral timescale. If we focus on the 
late stages of evolution, which are key from the gravitational-wave observation perspective, then this timescale would be of the order of a few minutes. 
If the  involved reactions are slower than this, then a perturbed fluid element must retain its identity (composition) through the evolution. 
We expect this limit to be relevant for binary neutron stars, c.f. \cite{RG92}.

Let us first analyze the case where the relevant reactions are much faster than the tidal dynamics. In this case it is natural to consider an equation of state such that the pressure depends on two parameters; the energy density $\rho$ and a parameter $\beta$ that represents the deviation from chemical equilibrium. Since  
the fluid elements will have time to equilibriate with the surrounding fluid, they will remain at (local) chemical equilibrium and we 
will have $\Delta \beta = 0$. However, $\beta=0$ also in the background, which means that
\be
\delta \beta = \Delta \beta - \xi^a \nabla_a \beta = 0 \ , 
\ee
as well. The upshot of this is that the system behaves as a barotrope. The fast interactions do not affect the tidal perturbations, and cannot have an impact on the Love number. 

In the opposite case, when reactions are very slow, it is natural to consider an equation of state with two charge neutral components, neutrons (with index n)  and a conglomerate of protons and electrons (with index p), such that $P=P(n,x_\p)$ where $n=n_\n+n_\p$ is the total baryon number density and $x_\p= n_\p/n$ is the proton fraction. Assuming that each species is conserved, i.e., that the relevant reactions are too slow to equilibrate the matter,
but the two components are locked together (i.e. there is only one displacement vector \cite{RELLAG}), we replace \eqref{eq:deln2} by 
\beq
\Delta n_\x =  - \frac12 n_\x \perp^{ab} \Delta g_{ab} \ , \qquad \x=\p,\n \ .
\label{deln}\eeq
Using these relations, we find that  $\Delta n$ is still given by \eqref{eq:deln2} and we also see that $\Delta x_\p = 0$.
These results simply represent the assumption of frozen composition. It also follows that 
\be
\Delta P = \left( {\frac{\partial P}{\partial n}} \right)_{x_\p} \Delta n \ , \quad \mbox{and} \quad 
\Delta \rho = \left( {\frac{\partial \rho}{\partial n} } \right)_{x_\p} \Delta
n \ . 
\ee
That is, we have
\be
\Delta P =  \left( {\frac{\partial P}{\partial \rho} }\right)_{x_\p} \Delta \rho \ . 
\ee
In other words, the Eulerian variations are related by 
\be
\delta P = \left( {\frac{\partial P}{\partial \rho} } \right)_{x_\p} \delta \rho + A_a \xi^a  \ ,
\label{dp_strat}\ee
where
\be
A_a =\nabla_a P =   \left[ \left( {\frac{\partial p}{\partial \rho} } \right)_{x_\p} - 
 \left({\frac{\partial p}{\partial \rho} } \right)_{\beta}  \right]\nabla_a \rho \ .\label{eq:SchwarzDisc}
\ee
where $\beta$ has the same meaning as in the case of fast reactions. The magnitude of $A^a$ defines the 
 Schwarzschild discriminant \cite{RG92}.

Let us now consider the perturbed Einstein equations in the case of  frozen composition. Since we made no assumptions about $\delta P$ and $\delta \rho$ when we discussed the perturbation equations, the analysis in Section~\ref{sec:static_perfect} must remain unchanged. In particular, the final differential equation for $H_0$ remains unaltered. However, the fact that this equation can be derived in two different ways is now important. As we have already mentioned, the problem is overdetermined unless \eqref{prhorel} holds. 
Combining this condition with  \eqref{dp_strat} we see that we must have
\be
\left[ \left( \frac{\partial P}{ \partial \rho } \right)_{x_\p} - 
 \left(\frac{\partial P}{ \partial \rho } \right)_{\beta}  \right] \Delta \rho = 0 \ , \quad \longrightarrow \quad \Delta \rho = 0 \ .
\ee
That is, the Lagrangian variation of the density must vanish identically. 
Moreover, we are led to the (possibly surprising) conclusion that {\em internal stratification has no effect on the Love number}.

These arguments suggest that Love number measurements will only reveal limited information about the equation of state. We may be able to constrain the supranuclear equation of state, e.g. in terms of the inferred compactness, but not probe the detailed composition. 

\subsection{Numerical results: The crust}

In order to quantify the role of the crust elasticity, we need to study the problem numerically for specific neutron stars models. In this first proof-of-principle study we will consider a simple model, based on a polytropic equation of state and an analytic model for the crust's shear modulus. In principle, we are ready to implement more realistic descriptions of both the bulk behaviour and the elasticity, but before doing this it makes sense to develop the computational strategy for this simpler model problem. An obvious advantage of the simplified description is that it is relatively straightforward to write down a set of non-dimensionalised perturbation equations, see Appendices~\ref{app:DimLess} and \ref{App:DimLessCrust}.

\begin{figure}[h]
\centering
\includegraphics[height=8cm,clip]{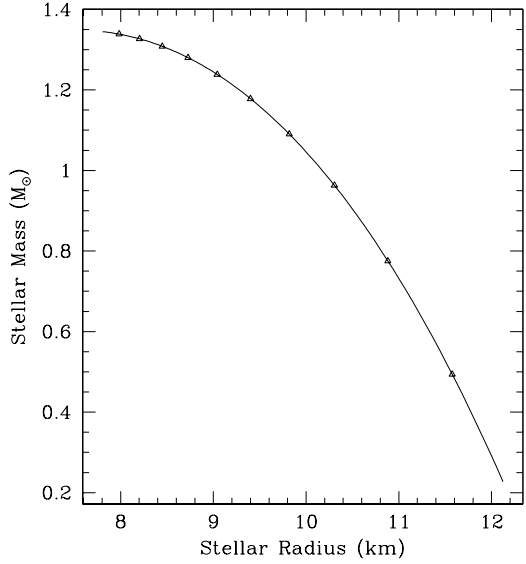}
\caption{The mass-radius curve for the stellar models considered in Table~\ref{table:1} (represented by triangles), showing that the considered models are stable to radial perturbations. In comparing to the data in the Table it is useful to recall that $1M_\odot = 1.4773\ \mathrm{km}$ in geometric units.}
\label{mvr}
\end{figure}

In order to facilitate a direct comparison to previous work \cite{Hinderer:2008,Poisson:2009,Damour:2009} we will assume that 
matter is described by the polytropic equation of state,
\be
 P = K \rho^{1+1/N} \ ,  \label{eq:EoS} 
\ee
where $K$ and $N$ are constant. In addition, we need to specify the shear modulus $\check{\mu}$.
The outer regions of an astrophysical neutron star are composed of i) a thin fluid ocean, below the density at which the crust melts for the given temperature (typically, $\rho < 10^7 \, \text{g cm}^{-3}$), ii) the outer crust, reaching up  to neutron drip  ($10^7\, \text{g cm}^{-3} \le \rho \le 10^{11}\, \text{g cm}^{-3}$), and iii) the inner crust, where superfluid neutrons permeate the nuclear lattice ($10^{11}\, \text{g cm}^{-3} < \rho < 10^{14}\, \text{g cm}^{-3}$)
Moreover, the bottom layers of the crust may be in the so-called pasta phase with rather different elastic properties \cite{LivRevCrust,Watanabe:2000}. We will not consider this possibility here. 
Our simplified model has a single elastic crust of thickness $\Delta R_c$. 
We will present results for a set of   (moderately realistic) polytropic models  for which  the crust region starts at $2\times10^{14}\ \mathrm{g/cm}^3$, and stops at $10^7\ \mathrm{g/cm}^3$. The non-crust regions are treated as perfect fluids. 
In the crust we implement a simple linear shear modulus \cite{LivRevCrust},
\be
 \muc = \kappa P + \mu_0,\label{eq:shearmu}
\ee
where $\kappa$ is a scaling constant and $\mu_0$ is a constant allowing us to consider an adjustable shear modulus at the top of the crust. Both $\kappa$ and $\mu_0$ are tuneable parameters.  The equations presented in Sec.~\ref{sec:crust_eqns} are, of course, independent of the specific form of the shear modulus \eqref{eq:shearmu}. Hence, the implementation of more realistic models would be straightforward. 

%

\begin{figure}[h]
\centering
\includegraphics[height=8cm,clip]{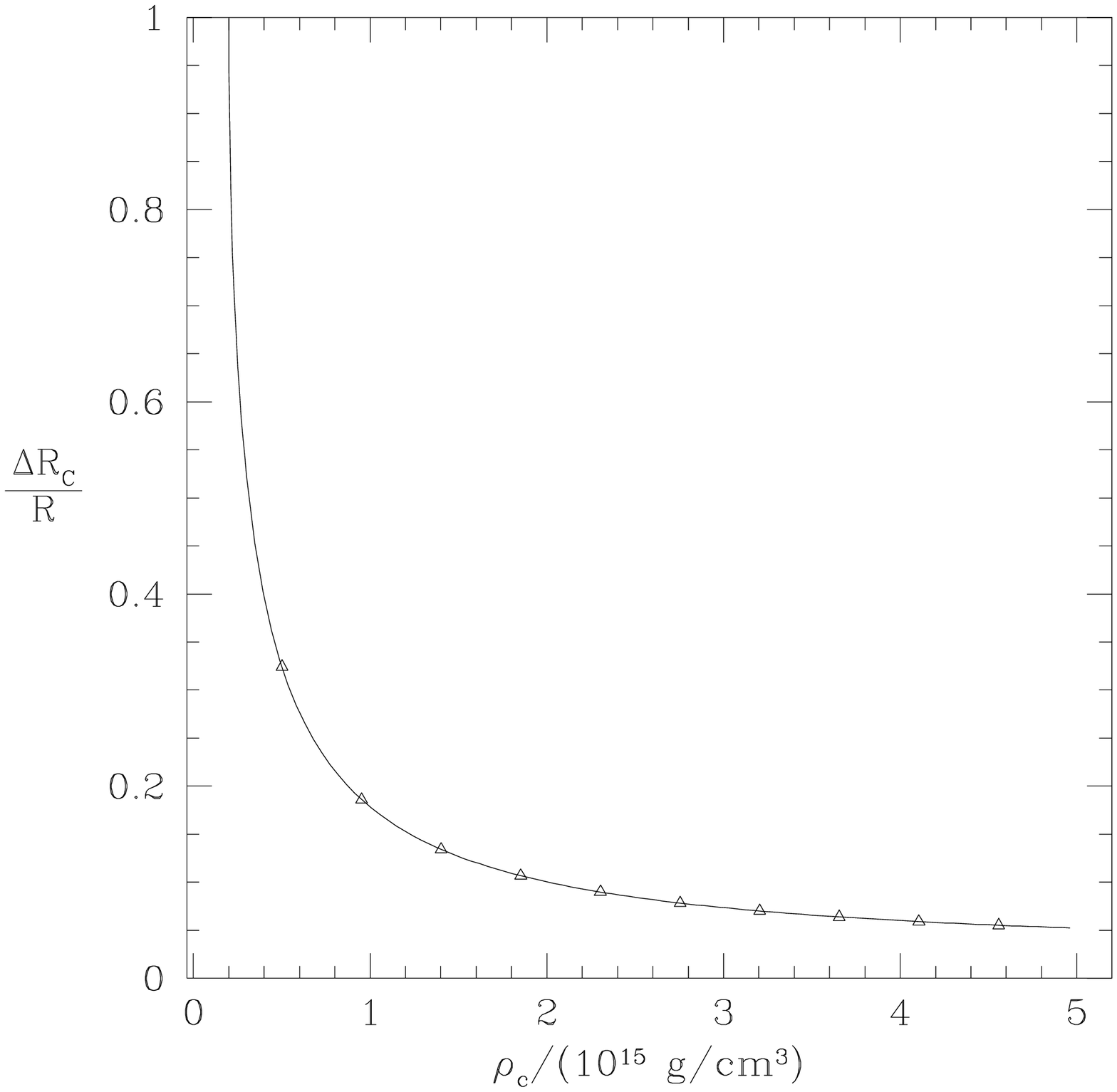}
\includegraphics[height=8cm,clip]{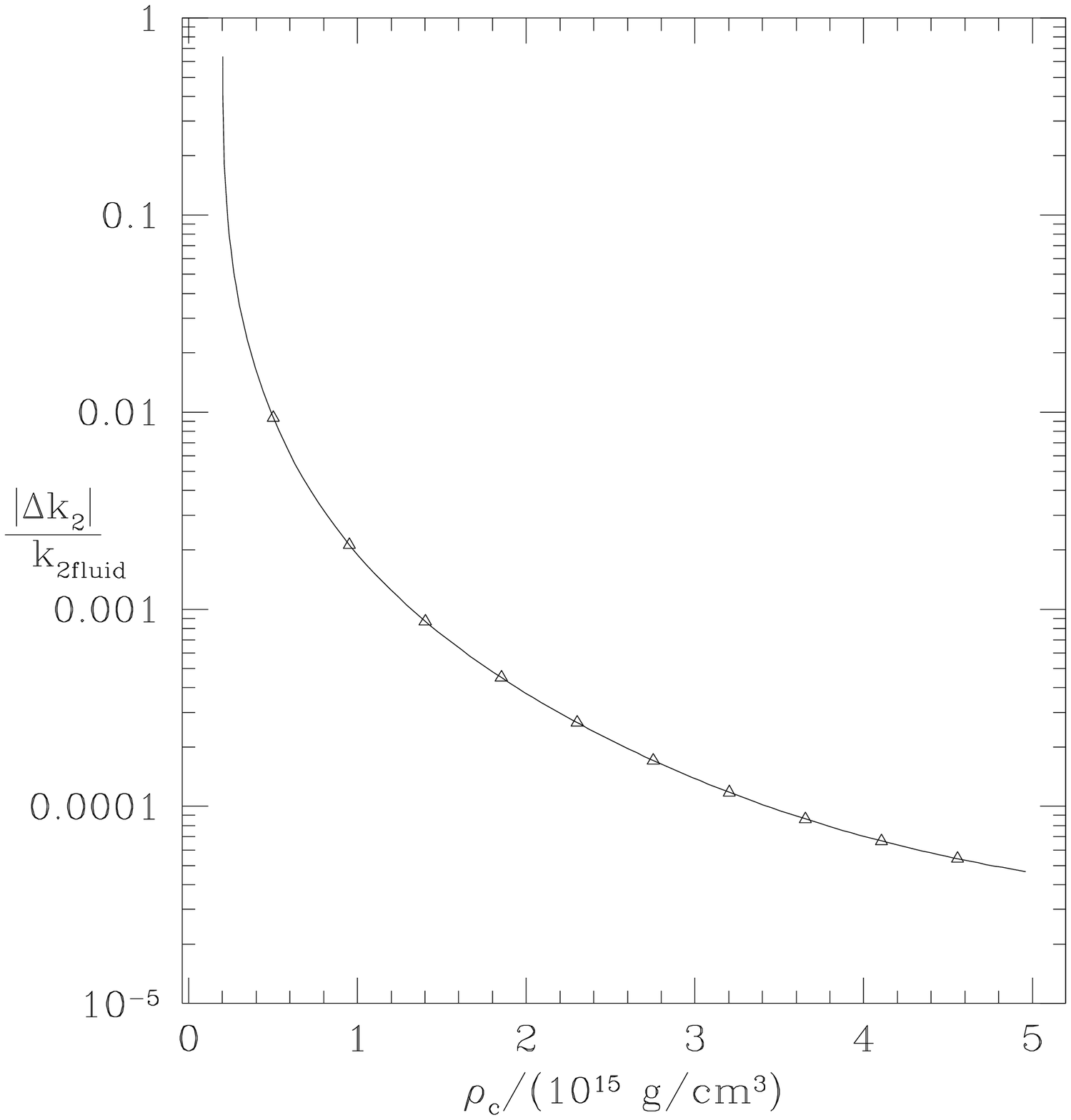}
\caption{Left panel: The ratio of crust thickness to stellar radius as function of the central density. The results show that the  crust occupies a much larger fraction of lower density stars. Right panel:  The relative change in Love number due to the presence of the crust as function of the central density. As the central density approaches the crust-core transition density, $2\times10^{14}\;\rm{g/cm}^3$, the contribution to the Love number becomes larger due to the fact that the crust occupies a much larger fraction of the star. Triangles correspond to the models listed in Table~\ref{table:1}.}
\label{dk2fig}
\end{figure}

We have, first of all, tested our numerical implementation by comparing the results for a fluid model to those of \cite{Poisson:2009} and \cite{Hinderer:2008}, achieving 
good agreement and hence confirming that  
the Love number is smaller for  larger values of the polytropic index. The tidal response is weaker for stars that are more centrally condensed, which is natural.
Adding the crust to these models, we obtain the results given in Table \ref{table:1}. The results were obtained for an $N=1$ polytrope with $K=100\ \mathrm{km}^2$  and the crust model \eqref{eq:shearmu}. The numerical results given in Table~\ref{table:1} were obtained for  a shear modulus $\kappa = 0.015$ and a shear constant $\mu_0 = 3 \times 10^{-14}\ \mathrm{km}^{-2}$, but we have confirmed the general behaviour 
by varying these parameters.
The chosen models are all stable to radial perturbations, c.f., the mass-radius curve in Figure~\ref{mvr}. As expected, the presence of the crust has only a small impact on the tidal Love number. 
By comparing the relative change in the Love number with increasing central density to the corresponding change in crust thickness (the left and right panels of Figure~\ref{dk2fig}, respectively), we see that the effect increases as more of the star becomes elastic. 
This is, of course, as expected. 

The main message is that we have successfully implemented the crust, and the numerical solution is accurate enough to distinguish the small effect that the elasticity has in this problem. The qualitative behaviour of the results is easily explained. 
The results in Table~\ref{table:1} show that presence of the crust decreases the tidal Love number. This is expected, since the Love number is a measure of the response to the presence of an external gravitational field. In a fluid model we expect a larger distortion since fluids do not sustain shear stresses. Crustal models are able to resist deformations, and are thus expected to have smaller Love numbers.

\begin{table}
\begin{center} {\tiny
\begin{tabular}{|p{0.7in}|p{0.75in}|p{0.75in}|p{0.75in}|p{0.6in}|p{0.7in}|p{0.7in}|p{0.7in}|}
\hline
$\rho_c$ $(\rm{km}^{-2})$ &  $C$ & $k_{2\;\rm{crust}}$ & $k_{2\;\rm{fluid}}$ & $\Delta k_{2}/k_{2\;\rm{fluid}}$ & $\Delta R_c$ (km) & $R$ (km) & $M$  (km)\\
\hline
3.378893e-03  & 0.24813001129(7) & 0.02413726613(6) & 0.024138574(6) & -5.420(7)e-05 & 0.4412458028(3) & 7.9838719798(7) & 1.9810382445(6)\\ 
3.044717e-03  & 0.23938571005(1) & 0.02748474657(2) & 0.027486580(6) & -6.672(4)e-05 & 0.4847246944(3) & 8.2035091998(7) & 1.9638028747(2)\\
2.710540e-03  & 0.22915622628(4) & 0.03172353529(9) & 0.031726271(2) & -8.623(6)e-05 & 0.5398838957(1) & 8.4494607108(1) & 1.9362465306(3)\\
2.376364e-03 & 0.21709115810(1) & 0.03719492461(6) & 0.037199305(6) & -1.177(7)e-04 & 0.6117312291(5) & 8.7266756593(2) & 1.8944841252(5)\\
2.042188e-03  & 0.20272655170(3) & 0.04441587924(3) & 0.044423470(8) & -1.708(9)e-04 & 0.7085096707(6) & 9.0414350438(4) & 1.8329389488(8)\\
1.708012e-03 & 0.18543653698(4) & 0.05419576631(7) & 0.054210160(9) & -2.655(3)e-04 & 0.8447708420(0) & 9.4018321596(9) & 1.7434431969(9)\\
1.373836e-03  & 0.16435873698(6) & 0.06785042152(8) & 0.067880961(4) & -4.499(0)e-04 & 1.0487697824(5) & 9.8184762317(9) & 1.6137523525(9)\\
1.039659e-03  & 0.13827570668(3) & 0.08761155990(9) & 0.087687716(3) & -8.684(9)e-04 & 1.3835220344(5) & 10.305549472(1) & 1.4250071360(2)\\
7.054831e-04  & 0.10541985896(6) & 0.11742663334(5) & 0.117676862(9) & -2.126(4)e-03 & 2.0250735506(0) & 10.882441793(5) & 1.1472254790(8)\\
3.713069e-04  & 0.06313942900(5) & 0.16404412956(2) & 0.165596266(2) & -9.373(0)e-03 & 3.7551538607(9) & 11.576364548(1) & 0.7309250475(2)\\
\hline
\end{tabular}
}
\caption{A sample of numerical results comparing the Love number for fluid models and the elastic crust models developed in this work. 
The results are obtained using an $N=1$ polytrope with $K=100\ \mathrm{km}^2$.
The stellar models are determined by the central density $\rho_{\rm{c}}$. We provide the resulting compactness, $C=M/R$, mass, $M$, radius, $R$, and crustal thickness, $\Delta R_c$, for the background star. The  tidal Love number for both the crust, $k_{2\;\rm{crust}}$, and purely fluid models, $k_{2\;\rm{fluid}}$, are shown. From the differences between the final Love numbers, $\Delta k_2 = k_{2\rm{crust}}-k_{2\rm{fluid}}$, we see that the crust produces a very small correction to the tidal Love number. For this table we used the elastic parameters $\kappa=0.015$ and $\mu_0=10^{-14}\ \mathrm{km}^{-2}$.}
\label{table:1}
\end{center}
\end{table}


\section{Conclusions}\label{sec:conclusion}

We have extended the discussion of tidally deformed relativistic stars by including the effects of internal composition stratification and the presence of an elastic crust,  relevant for mature neutron stars. The most important development concerns the formalism required to account for more realistic neutron star physics. Building on a recent extension \cite{RELLAG}
of relativistic Lagrangian perturbation theory to the multi-fluid setting (allowing for one of the components to be elastic), 
we have formulated the tidal deformation problem for realistic neutron star models. The final model is obviously still ``incomplete", as we are ignoring the magnetic field and we are not accounting for superfluidity  (which should be relevant both in the crust and the fluid core), but it nevertheless allows us to consider astrophysically relevant questions. 

This paper should be seen as a first, proof-of-principle, study of the relevant issues. We have shown that (perhaps unexpectedly) the tidal 
deformations are not affected at all by composition variations in the star's core. Having implemented the formalism required to account for the crust elasticity, and solved the problem numerically for a simple model problem (based on a polytropic equation of state and a simplistic model for the 
crust shear modulus), we have also demonstrated that the presence of the crust has a (predictably) small effect on the tidal Love number. We have considered how this effect varies with the crust parameters, and confirmed that the results agree with intuition.

From a formal point-of-view, these are important developments, but it should be stressed that we do not expect the new features to leave an observable imprint on a binary gravitational-wave signal. This is fairly obvious since the Love number leaves an imprint that is barely detectable by future generations of gravitational-wave detectors in the first place \cite{Flanagan:2008} and \cite{READ}, and we are quantifying ``small'' corrections to it. Having said that, it is clear that we 
need to move beyond  back-of-the-envelope estimates in this problem area and develop the computational technology required to model 
real neutron stars and actual astrophysical scenarios.  The present work represents an important step towards this goal. 
By developing the required relativistic perturbation formalism for the tidal interaction problem, we lay the foundation for work on general crust asymmetries, e.g., neutron star ``mountains''  relevant for  gravitational-wave astronomy.  This problem has not yet been considered in general relativity, which is required if we want to use realistic equations of state. A closely related problem concerns the fracture of the crust during binary inspiral, the effect that this may have on a detected gravitational-wave signal and possible associated electromagnetic signatures. Another important problem concerns the quasiperiodic oscillations observed in the tails of magnetar flares \cite{Israel, Watts:2007,Samuelsson:2009}. In this context, a key issue concerns the build-up of stresses in the crust of these strongly magnetized stars, the eventual fracture and the induced dynamics of the coupled magneto-elastic system.  These are exciting problem of immediate astrophysical interest, and we hope to make progress on them in the near future.

\section*{Acknowledgements}

AJP acknowledges support from the Virgo-EGO Scientific Forum in terms of a fellowship.
NA, IH and DIJ acknowledge support from STFC via grant number ST/H002359/1. 
We  also acknowledge support from COMPSTAR (an ESF Research Networking Programme).

\appendix

\section{Appendix: The non-dimensionalised problem}

\subsection{Dimensionless Background Problem}\label{app:DimLess}

The equation of state \eqref{eq:EoS} allows us to write the equilibrium equations, \eqref{eq:dMdr} and \eqref{eq:dPdr} in a dimensionless form, c.f.,  \cite{Tooper:1964,Poisson:2009}, 
\begin{align}
 \rho &= \rho_{\rm{c}} \vartheta^N \ ,\\
 P &= P_{\rm{c}} \vartheta^{N+1} \ ,\\
 m &= m_0\eta \ ,\\
 r &= r_0\zeta \ ,
\end{align}
where $\rho_c$ is the density at the centre of the star, and $P_c = K\rho_{\rm{c}}^{1+1/N}$ is the central pressure. The units of mass become $m_0=4\pi r_0^3\rho_{\rm{c}}$, and radius, $r_0^2 = \frac{(N+1)b}{4\pi\rho_{\rm{c}}}$ where we define $b=P_{\rm{c}}/\rho_{\rm{c}}$. As discussed in \cite{Poisson:2009}, $b$ is a parameter that may be used to gauge the relativistic behaviour of a model. In the limit $b\rightarrow0$ the model becomes Newtonian.

Using the dimensionless variables the background equations \eqref{eq:dMdr} and \eqref{eq:dPdr} are re-written as \cite{Poisson:2009},
\begin{align}
\dot{\eta} &= \zeta^2\vartheta^N,\\
\dot{\vartheta} & = -\frac{\left(\eta+b\zeta^3\vartheta^{N+1}\right)\left(1+b\vartheta\right)}{\zeta^2 f} \ , 
\end{align}
where $f = 1-2(N+1)b\eta/\zeta$. Using these variables and considering regularity at the centre, we have the conditions $\vartheta_{\zeta=0}=1$, $\mu_{\zeta=0}=0$ \cite{Poisson:2009}.

For numerical accuracy we change variables one last time. We use $\chi=\eta/\zeta^3$ and $x=\ln{\zeta}$ \cite{Poisson:2009}. By using these coordinates, we allow for greater accuracy near the centre of the star. The final form of the background equations becomes
\begin{align}
 \dot{\chi} &= \vartheta^N-2\chi \ ,\\
 \dot{\vartheta} &= -\frac{\zeta^2}{f}\left(\chi+b\vartheta^{N+1}\right)(1+b\vartheta) \ , 
\end{align}
where $f=1-2(n+1)b\zeta^2\chi$. Due to difficulties integrating from the actual origin of the star, we start the integration at $x=-30$ which corresponds to $\zeta \sim 10^{-14}$ . The integration terminates at $\zeta=\zeta_f$ where $\vartheta$ drops below a predetermined numerical tolerance. With these definitions we define the mass, $M$, and radius, $R$, of the star to be \cite{Poisson:2009}
\begin{align}
 M &= \frac{(N+1)^{3/2}K^{N/2}}{\sqrt{4\pi}} b^{(3-N)/2}\zeta_f^3\chi_f \ ,\\
 R &= \sqrt{\frac{N+1}{4\pi}}K^{N/2} b^{(1-N)/2}\zeta_f,
\end{align}
where $\chi_f = \chi(\zeta_f)$.

We note that \cite{Poisson:2009} chose to use $b$ in place of the central density $\rho_{\rm{c}}$ to label a stellar model. This is possible since $\rho_{\rm{c}}=b^N/K^N$ where they use $(K,N)$ to parameterize their equation of state. Motivated by the need to determine a physical location of the crust, we choose to parameterize our equation of state using $(\rho_{\rm{c}},N)$ using $K = b/\rho_{\rm{c}}^{1/N}$. Thus we can specify the central density, as well as the crust-core  and  crust-ocean transition densities. This choice of parameterization does not impact the form of the stellar compactness or the fluid tidal Love number as defined in \cite{Poisson:2009}.

\subsection{Dimensionless Crustal Perturbation Equations}\label{App:DimLessCrust}
We perform the same procedure used in Sec.~\ref{app:DimLess} to reduce our problem to a dimensionless set of equations. First we note that our perturbation $\xi^a$ has units of distance, so we define
$V \rightarrow r_0^2\tilde{V}$ and $W \rightarrow r_0^2\tilde{W}$. We also note that the shear modulus $\muc$ has units of pressure, so we define $\muc\rightarrow\rho_c\tilde{\mu}$. The factor $b$ which should be present is absorbed into our definition of $\tilde{\mu}$.

Using this information in our definition of the traction reveals
\begin{align}
 T_1 &= \frac{4\tilde{\mu}}{3}\rho_{\rm{c}} r_0^2\left[\zeta^2(K-H_2)-l(l+1)\tilde{V}-2\zeta\dot{\tilde{W}}+(4-\zeta\dot{\lambda})\tilde{W}\right] \ ,\\
 \tilde{T}_2 &=-4\tilde{\mu}\rho_{\rm{c}} r_0^2(\zeta\dot{\tilde{V}}-2\tilde{V}+e^{\lambda}\tilde{W}) \ , 
\end{align}
where the  dots denotes differentiation with respect to $\zeta$. For notational simplicity we also define $\delta \tilde{P}=r_0^2\delta P$.

This gives us the dimensionless form of the crustal equations;
\begin{align}
\dot{\tilde{W}} & = \left(\frac{2}{\zeta} - \frac{\dot{\lambda}}{2}\right)\tilde{W} + \frac{\zeta(K-H_0)}{2} -8\tilde{\mu}(N+1)b\zeta\tilde{V} -\frac{l(l+1)}{2\zeta}\tilde{V}- \frac{3T_1}{8\tilde{\mu}\zeta}\frac{4\pi}{(N+1)b} \ , \\
\dot{\tilde{V}} &= \frac{2\tilde{V}}{\zeta} - \frac{e^{\lambda}}{\zeta}\tilde{W} - \frac{T_2}{4\tilde{\mu}\zeta}\frac{4\pi}{(N+1)b}\ , \\
\dot{K} &= \dot{\nu}H + \dot{H} + \frac{8\tilde{\mu}}{\zeta}(\zeta\dot{\nu}+2)\tilde{V}(N+1)b -8\pi\frac{T_2}{\zeta}\ , \\
\dot{T}_2 &= \left[\frac{\dot{\lambda}-\dot{\nu}}{2}-\frac{1}{\zeta}\right]T_2-2e^{\lambda}\zeta \delta \tilde{P}+\frac{1}{8\pi}(\dot{\nu}+\dot{\lambda})H+\frac{4\tilde{\mu}}{\zeta}e^{\lambda}[2-l(l+1)]\tilde{V}\frac{4\pi}{(N+1)b}+\frac{e^{\lambda}T_1}{\zeta}\ , \\
\ddot{H} &+ \left[\frac{2}{\zeta}+\frac{\dot{\nu}-\dot{\lambda}}{2}\right]\dot{H}+\left[\frac{-l(l+1)e^{\lambda}}{\zeta^2}+\frac{2(1-e^{\lambda})}{\zeta^2}+\frac{\dot{\lambda}+3\dot{\nu}}{\zeta}-\dot{\nu}^2\right]H \nonumber\\
&=-8\pi\left[e^{\lambda}(3+1/c_{\rm{s}}^2)\delta \tilde{P}+ 16\tilde{\mu}\frac{(N+1)b}{4\pi}\left[1-e^{\lambda}+\zeta(\dot{\nu}+\tfrac{1}{2}\dot{\lambda})-\frac{\zeta^2\dot{\nu}^2}{4}\right]\frac{\tilde{V}}{\zeta^2} +4\dot{\nu}\frac{(N+1)b}{4\pi}\dot{\left(\tilde{\mu}\tilde{V}\right)}\;+\frac{\dot{\nu}T_2}{\zeta}\right] \ . 
\end{align}
Here we used $\rho_{\rm{c}} r_0^2 = (N+1)b/4\pi$.
We also re-write the algebraic relations for $T_1$ \eqref{alg1} and $\delta P$ \eqref{alg2} as
\begin{align}
 T_1 &= \frac{(N+1)b}{4\pi}\frac{8\tilde{\mu}}{3}\left(\frac{4\pi \zeta^2}{(b\vartheta+1)\vartheta^n (n+1)bc^2_{\rm{s}}}\delta \tilde{P}+\frac32\zeta^2K-\frac32l(l+1)\tilde{V}+\left(3-\frac{\zeta\dot{\nu}}{2c_{\rm{s}}^2}\right)\tilde{W}\right)\label{eq:T1} \ ,\\
 16\pi e^{\lambda}\zeta^2\delta \tilde{P} &= \zeta^2\dot{\nu}\dot{H}+[2-l(l+1)]e^{\lambda}K+[l(l+1)e^{\lambda}-2+\zeta^2\dot{\nu}^2]H\nonumber\\
&+8(N+1)b\tilde{\mu}[\zeta^2\dot{\nu}^2+[l(l+1)-2](1-e^{\lambda})]\tilde{V}-8\pi(\zeta\dot{\nu}+2)T_2+8\pi(e^{\lambda}-3)T_1\label{eq:vP} \ .
\end{align}

Using the same computational variables as in \cite{Poisson:2009} for the crust, we eliminate division by the radial coordinate. This is more useful for calculations that include the centre of the star; however, it is a simple coordinate transform that allows us to join the fluid core to the crust without changing coordinates.
\begin{align}
\partial_x{\tilde{W}} & = \left(2 - \frac{\partial_x{\lambda}}{2}\right)\tilde{W} + \frac{e^{-2x}(K-H_0)}{2} -8\tilde{\mu}(N+1)b e^{-2x}\tilde{V} -\frac{l(l+1)}{2}\tilde{V}- \frac{3T_1}{8\tilde{\mu}}\frac{4\pi}{(N+1)b}\ , \label{eq:dimless_W} \\
\partial_x{\tilde{V}} &= 2\tilde{V} - e^{\lambda}\tilde{W} - \frac{T_2}{4\tilde{\mu}}\frac{4\pi}{(N+1)b} \ ,\label{eq:dimless_V} \\
\partial_x{K} &= (\partial_x\nu)H + \partial_x{H} + 8\tilde{\mu}(\partial_x\nu+2)\tilde{V}(N+1)b -8\pi T_2 \ , \\
\partial_x{T}_2 &= \left[\frac{(\partial_x{\lambda}-\partial_x{\nu})}{2}-1\right]T_2-2e^{\lambda}e^{-2x} \delta \tilde{P}+\frac{1}{8\pi}(\partial_x{\nu}+\partial_x{\lambda})H+4\tilde{\mu}e^{\lambda}[2-l(l+1)]\tilde{V}\frac{4\pi}{(N+1)b}+e^{\lambda}T_1\ , \\
\partial_x^2{H} &+ \left[1+\frac{(\partial_x{\nu}-\partial_x{\lambda})}{2}\right]\partial_x{H}+\left[-l(l+1)e^{\lambda}+2(1 - e^{\lambda})+(\partial_x{\lambda}+3\partial_x{\nu})-(\partial_x\nu)^2\right]H \nonumber\\
&=-8\pi\left[e^{\lambda}(3+1/c_{\rm{s}}^2)e^{-2x}\delta \tilde{P}\right.\nonumber\\
&+\left. 16\tilde{\mu}\frac{(N+1)b}{4\pi}\left[1-e^{\lambda}+(\partial_x{\nu}+\tfrac{1}{2}\partial_x{\lambda})-\frac{\partial_x{\nu}^2}{4}\right]\tilde{V}+4\partial_x{\nu}\frac{(N+1)b}{4\pi}\partial_x{\left(\tilde{\mu}\tilde{V}\right)}+\partial_x{\nu}T_2\right] \ .
\end{align}
The algebraic relations are also slightly modified to
\begin{align}
 T_1 &= \frac{(N+1)b}{4\pi}\frac{8\tilde{\mu}}{3}\left(\frac{4\pi e^{-2x}}{(b\vartheta+1)\vartheta^n (N+1)bc^2_{\rm{s}}}\delta \tilde{P}+\frac32e^{-2x}K-\frac32l(l+1)\tilde{V}+\left(3-\frac{\partial_x{\nu}}{2c_{\rm{s}}^2}\right)\tilde{W}\right)\label{eq:nT1}\ , \\
 16\pi e^{\lambda}e^{-2x}\delta \tilde{P} &= \partial_x{\nu}\partial_x{H}+[2-l(l+1)]e^{\lambda}K+[l(l+1)e^{\lambda}-2+(\partial_x\nu)^2]H\nonumber\\
&+8(N+1)b\tilde{\mu}[(\partial_x\nu)^2+[l(l+1)-2](1-e^{\lambda})]\tilde{V}-8\pi(\partial_x\nu+2)T_2+8\pi(e^{\lambda}-3)T_1\label{eq:nvP} \ .
\end{align}

We note that, with our particular formulation it is not obvious that we may consider the Newtonian limit since \eqref{eq:dimless_W} and \eqref{eq:dimless_V} are divergent in that limit. To obtain a non-divergent form we would need to use the equations that follow from the divergence of the stress-energy. Since the Newtonian limit is not the focus of our study we do not pursue this form of the equations.  We also note that the given dimensionless analysis is only valid for a polytropic equation of state.

\bibliographystyle{plainnat}

\end{document}